\documentclass[12pt,preprint]{aastex} 

\shorttitle{PO/LCO Atlas of BCDs: II. Surface Photometry}
\shortauthors{Gil de Paz et al.}

\begin{document}

\title{Palomar/Las Campanas Imaging Atlas of Blue Compact Dwarf
Galaxies: II. Surface Photometry and the Properties of the Underlying 
Stellar Population}

\author{A. Gil de Paz\altaffilmark{1}, and B. F. Madore\altaffilmark{1,2}}

\altaffiltext{1} {The Observatories, Carnegie Institution of
Washington, 813 Santa Barbara Street, Pasadena, CA 91101; agpaz@ociw.edu}
\altaffiltext{2} {NASA/IPAC Extragalactic Database, California
Institute of Technology, MS 100-22, Pasadena, CA 91125;
barry@ipac.caltech.edu} 

\begin{abstract}
We present the results from an analysis of surface photometry of $B$,
$R$, and H$\alpha$ images of a total of 114 nearby galaxies
($v_{\mathrm{helio}}$$<$4000\,km\,s$^{-1}$) drawn from the Palomar/Las
Campanas Imaging Atlas of Blue Compact Dwarf galaxies. Surface
brightness and color profiles for the complete sample have been
obtained. We determine the exponential and S\'ersic profiles that best
fit the surface brightness distribution of the underlying stellar
population detected in these galaxies. We also compute the ($B-R$)
color and total absolute magnitude of the underlying stellar
population and compared them to the integrated properties of the
galaxies in the sample. Our analysis shows that the ($B-R$) color of
the underlying population is systematically redder than the integrated
color, except in those galaxies where the integrated colors are
strongly contaminated by line and nebular-continuum emission. We also
find that galaxies with relatively red underlying stellar populations
(typically ($B-R$)$\geq$1\,mag) show structural properties compatible
with those of dwarf elliptical galaxies (i.e$.$ a smooth light
distribution, fainter extrapolated central surface brightness and
larger scale lengths than BCD galaxies with blue underlying stellar
populations). At least $\sim$15\% of the galaxies in the sample are
compatible with being dwarf elliptical (dE) galaxies experiencing a
burst of star formation. For the remaining BCD galaxies in the sample
we do not find any correlation between the recent star formation
activity and their structural differences with respect to other types
of dwarf galaxies.
\end{abstract}


\section{Introduction}
\label{intro}

The vast majority of the Blue Compact Dwarf (BCD) galaxies are
currently forming stars at a very high rate, as suggested by the large
fraction of them ($>$93\%; Gil de Paz, Madore, \& Pevunova 2003)
showing significant H$\alpha$ emission (EW$>$20\,\AA). The star
formation rate (SFR) derived for these galaxies ranges up
$\sim$10\,M$_{\odot}$\,yr$^{-1}$ (Fanelli et al$.$ 1988; Gil de Paz et
al$.$ 2003). This relatively high SFR, combined with a moderate gas
content, implies gas-consumption time-scales of $\sim$10$^{9}$\,yr,
which are much shorter than the age of the Universe.

This high SFR, combined with the low present-day metal abundances (see
e.g$.$ Hunter \& Hoffman 1999), led Searle et al$.$ (1973) to suggest
that either these objects are intrinsically young galaxies or that
they have had an episodic star formation history involving very short
bursts of star formation followed by long quiescent periods. It is now
widely accepted that most of these objects possess a relatively
evolved underlying stellar population (USP hereafter) associated with
their low-surface-brightness (LSB) envelopes, i.e$.$ they are not
exclusively young galaxies (Schulte-Ladbeck et al$.$ 1999, 2001;
Drozdovsky et al$.$ 2001; Crone et al$.$ 2000, 2002; Gil de Paz et
al$.$ 2000a,b; Papaderos et al$.$ 1996a,b). Despite considerable
recent progress, the properties of this USP are still poorly
known. Although the presence of an evolved population supports a
relatively episodic star formation history it has been recently shown
that these episodes of star formation may, in fact, last as long as
10$^{8}$\,yr (Crone et al$.$ 2002; Papaderos et al$.$ 2002). However,
it is still necessary that the SFR in BCDs had been much lower in the
past (and for extended periods of time) as compared to today. It has
been suggested that, even during the so-called quiescent phases, star
formation could have taken place at a very low level, at a rather
continuous rate (Legrand 2000).

The following questions naturally arise: What did BCD galaxies look
like during these long periods of time of low (or null) star-formation
activity? Were they similar to local dwarf elliptical galaxies today?
and, is there a single evolutionary model that can include all phases
and types of dwarf galaxies?

Current knowledge bearing on these questions comes mainly from the
analysis of the surface brightness profiles of the USP in BCDs and a
comparison of the profiles of dwarf irregular and dwarf elliptical
galaxies (Papaderos et al$.$ 1996; Noeske et al$.$ 2000; Cair\'os et
al$.$ 2003). These results seem to suggest that the LSB envelopes
associated with the USP of BCD galaxies are more compact and have a
higher central surface brightness than those seen in dI and dE
galaxies. Some additional clues have been recently provided by the
study of the dynamics of BCD neutral hydrogen halos (van Zee, Salzer,
\& Skillman 2001; Tajiri \& Kamaya 2002). These authors have shown
that the suggested evolution between BCDs and dE is difficult because
of the relatively low efficiency of stellar feedback in potentially
removing the HI envelopes from these galaxies. It is worth emphasizing
here that in all of these previous studies the number of objects under
consideration was quite limited and no definitive conclusions could be
drawn concerning existence or not of a unified evolutionary model for
dwarf galaxies.

In Paper~I (Gil de Paz et al$.$ 2003) we described the integrated
properties of a sample of 114 nearby galaxies in the Palomar/Las
Campanas Imaging Atlas of Blue Compact Dwarfs (BCD hereafter). In
this, the second paper of the series, we analyze the surface
brightness profiles of the sample in $B$ and $R$ bands and in the
light of H$\alpha$. The morphological information derived, along with
the properties of the USP detected in these galaxies, is now compared
with the integrated properties measured in Paper~I.

In Section~\ref{sampleobs} we briefly describe the sample and the
observations carried out within the Atlas. The procedures used to
derive the surface brightness and color profiles and the corresponding
best-fitting exponential and S\'ersic laws are described in
Sections~\ref{analysis.profiles} and \ref{analysis.fitting},
respectively. The analysis of the surface brightness profiles is
presented in Section~\ref{results.sbp}. We describe the structural
properties, color, and luminosity of the USP of these galaxies in
Section~\ref{results.lsb}. Finally, we discuss the implications of
this study regarding the existence of possible evolutionary links
between BCDs and dwarf elliptical galaxies
(Section~\ref{results.dwarfs}), and the possible impact of the recent
star formation on the structural properties of BCDs
(Section~\ref{results.sfh}). Our conclusions are summarized in
Section~\ref{conclusions}.

\section{Sample and observations}
\label{sampleobs}
The original sample described in Paper~I consisted of 114 galaxies. Of
these, 105 galaxies were finally classified as Blue Compact Dwarf
galaxies according to the set of quantitative criteria set out in that
paper. The criteria include considerations about the galaxy's peak
surface brightness, and the color at the position of this peak, along
with an upper limit in the galaxy integrated absolute $K$-band
luminosity (i.e., stellar mass). In this paper we have removed IC~10
from the sample of BCDs because of its very high Galactic extinction,
which makes the intrinsic luminosity and colors of the USP highly
uncertain. Although we computed the surface brightness profiles for
all the 114 galaxies, our conclusions are based exclusively on the
analysis of the properties of the final 104 BCD galaxies in the Atlas.

In Paper~I we also presented an extensive description of the
observations. Briefly, we observed 86 of the 114 Atlas galaxies at the
Palomar 60-inch telescope using a 2048$\times$2048 CCD in $B$, $R$,
and the appropriately redshifted H$\alpha$ narrow-band filter
($\Delta\lambda$$\sim$20\AA; see Paper~I for a description of the
filters used). Typical exposure times were 900\,s in $B$, 2700\,s in
$R$, and 5400\,s in H$\alpha$. The remaining 28 galaxies were observed
at the Las Campanas Observatory 100-inch (du Pont) telescope using a
similar 2048$\times$2048 CCD. In this case the exposure times were
900\,s in $B$ and $R$, and 1800\,s in narrow-band H$\alpha$
($\Delta\lambda$$\sim$65\AA).

\section{Analysis}
\label{analysis}

\subsection{Surface brightness and color profiles}
\label{analysis.profiles}

We used the flux-calibrated images of our Atlas to derive the surface
brightness profiles of the sample. We first interactively eliminated
foreground and background sources by interpolation using the same
criteria as in Paper~I (see also Gil de Paz et al$.$ 2000a). Then, we
fitted the galaxy isophotes using the iterative method described by
Jedrzejewski (1987) within the IRAF task {\sc ellipse}. The output of
this task provides the {\it equivalent radius} ($R^{*}=\sqrt{a \times
b}$) and mean intensity of the isophote, its rms uncertainty,
ellipticity, position angle, etc.

The isophotes were fitted using our $R$-band images (which were
typically deeper than the $B$-band exposures). Then, we computed the
mean flux and rms in the $B$-band and H$\alpha$ images using the
isophotes fitted in the $R$-band. In this way the colors were measured
in exactly the same regions. We started the fitting procedure at the
approximate position of the half-light radius, then we moved outward
from that position; and finally, we moved inward using constant linear
steps of between 1 and 5 pixels. The step size was determined by the
depth and quality (spatial resolution) of the image.

In order to compute the error in the surface brightness and color
profiles we start from the expression that relates the intensity in
counts per pixel with the surface brightness,
\begin{equation}
\mu_\lambda = C_\lambda - 2.5 \log ( I_\lambda -
I_{\mathrm{sky},\lambda}) + 5 \log (\mathrm{arcsec/pixel})
\label{mub}
\end{equation}
Thus, to a first approximation we can write the uncertainty in $\mu_\lambda$
as,
\begin{equation}
\Delta \mu_\lambda = \sqrt{\Delta C_\lambda^2 + \left({{2.5 \log(e)} \over {I_\lambda -
I_{\mathrm{sky},\lambda}}} \Delta ( I_\lambda -
I_{\mathrm{sky},\lambda} )\right)^2}
\label{dmub}
\end{equation}
That can be expressed in terms of the rms uncertainty along the isophote and the error on the sky level as,
\begin{equation}
\Delta \mu_\lambda = \sqrt{\Delta C_\lambda^2 + \left({{2.5 \log(e)} \over {I_\lambda -
I_{\mathrm{sky},\lambda}}}\right)^2 \left(\left({{\mathrm{rms}_{\mathrm{isophote}}}\over{\sqrt{N_{\mathrm{isophote}}}}}\right)^2 + \Delta I_{\mathrm{sky},\lambda}^2\right)}
\label{dmub2}
\end{equation}
The term $\Delta I_{\mathrm{sky},\lambda}^2$ is actually composed of
two terms: one is due to a combination of Poisson noise in the sky
values and high-frequency (pixel-to-pixel) flat-fielding errors, and
the other is due to low-frequency flat-fielding errors and the
presence of reflections or gradients in the image background. This
latter component may be dominant in the outermost part of the galaxy
profiles as shown by Noeske et al$.$ (2003). In order to determine
these numbers we measured the value of the sky and its standard
deviation in a total of 15-20 regions of $N_{\mathrm{region}}$ pixels
each around the position of the galaxy. If we now define
$<\sigma_{\mathrm{sky}}>$ and $\sigma^2_{<\mathrm{sky}>}$ as the mean
standard deviation and variance of the sky values measured in these
individual regions, respectively, we obtain,
\begin{equation}
 \Delta I_{\mathrm{sky},\lambda}^2 =
 {{<\sigma_{\mathrm{sky}}>^2}\over{N_{\mathrm{isophote}}}} +
 \max \left( \sigma_{\mathrm{<sky>}}^2 -
 {{<\sigma_{\mathrm{sky}}>^2}\over{N_{\mathrm{region}}}},0 \right)
\label{dmub3}
\end{equation}
The second term of the sum can be neglected if the low-frequency
flat-fielding errors are negligible compared with the combined effect
of the sky photon noise and the high-frequency flat-fielding errors.

In Figure~\ref{figure1} we plot the surface brightness and color
profiles for each of the 114 galaxies in the original Atlas
sample. The 1-$\sigma$ error bars plotted combine in quadrature the
standard deviation of the isophote mean and the error in the sky
subtraction. The calibration errors for each of the bands are
indicated by vertical bars in the lower-left corner of the
plot. Horizontal tick marks on the left vertical axis indicate the
value of the HWHM of the PSF. Figure~\ref{figure1} shows that all the
galaxies in the sample are clearly resolved, so these light profiles
can be adequately used to study the distribution of the stellar
populations in these galaxies. Even in the case of the very compact
galaxy UM~404, which was observed with a very poor seeing (almost
4\,arcsec in the $B$-band), its profile extends radially more than
three times the PSF HWHM. The surface brightnesses and colors have
been corrected for Galactic extinction but not for internal
extinction. Despite the very low metallicity and, therefore, low
expected dust content, internal extinction may be important in the
innermost regions, where most of the current star formation activity
is taking place (see Noeske et al$.$ 2003; Cannon et al$.$
2002). However, the effect of the dust on the structural properties
and colors of the USP derived in this paper, which are measured at a
considerable distance from the sites of current star formation, is
probably negligible.

The surface brightness profiles obtained show a very high central
surface brightness that decreases very rapidly with the galactocentric
radius. For many of the galaxies in our sample a radius can be
identified outside which the surface brightness starts to decrease at
a much slower rate following an approxiate exponential or de
Vaucoleurs $r^{1/4}$ law, depending on the galaxy. We interpret this
behavior as being due to the existence of two well-differentiated
stellar populations, a young population that would be responsible for
the high-surface-brighness (HSB) nuclear emission and a more evolved
(fainter and redder) underlying stellar population (USP) with a
smooth, low-surface-brightness profile dominating the outermost parts
of the galaxy's surface brightness profile. This idea, which had been
proposed in the past by different authors (see Papaderos et al$.$ 1996
and references therein), is supported by the fact than in many of the
galaxies in our sample this radius coincides with a flattening in the
color profile, also called the {\it transition radius} (Papaderos et
al$.$ 1996).

However, there are many cases where the analysis of the surface
brigthness profile alone does not allow identifying the radius from
which the USP starts to dominate the galaxy's light distribution. This
can be due to (1) the USP dominating the profile all the way to the
center of the galaxy, (2) the relative contribution in luminosity of
the two stellar populations changing gradually but in a very smooth
way across the galaxy, (3) the transition taking place at a
surface-brightness very close or below our detection limit (i.e. the
USP is undetected), or (4) the USP is not present, like it could the
case of those galaxies suspect of forming stars for the first time
(e.g. I~Zw~18; Tol~65). In Section~\ref{analysis.fitting} we will
describe the procedure used to determine (in a homogeneous way) the
radius outside which the galaxy's surface brightness profile is
dominated by the emission from its USP. 

\subsection{Profile fitting}
\label{analysis.fitting}

We did not find any simple function or combination of functions that
adequately reproduces the surface brightness profile of the galaxies
all the way from the nucleus to the outermost regions.
 
In order to compare the structural properties of the USP in our BCD
sample with those observed in other dwarf galaxies we fitted the
surface brightness of the USP using both an exponential and
a S\'ersic law. The free parameters for this fitting procedure where
those of the exponential and S\'ersic laws plus the position of the
radius from where the surface brightness profile is assumed to be
dominated by the USP. In those cases where there was an obvious
flattening of the color profile we only considered points external to
the {\it transition radius}. The best-fitting set of parameters was
obtained by minimizing the reduced $\chi^2$ (normalized to the degrees
of freedom) of the fit. No less than 5 points were used for each fit.

The position of the best-fitting innermost point of the fit (i.e. the
radius outside which the USP dominates the profile) is shown
in the Figure~\ref{figure1} by vertical tick marks at the bottom axis
of the surface-brightness plot. In most cases this radius is similar
for both the exponential and S\'ersic-law fits, which suggests the
presence of a clear change in the surface brightness profile at this
radius. In those cases where a significant difference is seen between
the two fits, the best-fitting innermost point for the exponential law
is usually placed at a larger galactocentric distance than the
corresponding for the S\'ersic law. In this plot we also show the
resulting best-fitting exponential and S\'ersic laws in the range of
the fit (solid lines) and the corresponding extrapolation toward the
galaxy center (dashed lines).

In Tables~\ref{table1} and \ref{table2} the parameters of the
best-fitting exponential and S\'ersic laws are given. In
Figures~\ref{figure2}a to \ref{figure2}f we show the parameters and
$\chi_r^2$ from the fit of an exponential law to the $B$-band
(\ref{figure2}a-\ref{figure2}c) and $R$-band
(\ref{figure2}d-\ref{figure2}f) profiles. The distributions of
best-fitting S\'ersic indices for the $B$ and $R$ profiles are shown
in panels \ref{figure2}g and \ref{figure2}h, respectively.

Tables~\ref{table1} and \ref{table2} show that the application of
either of these laws to the broad-band surface brightness profiles of
the USP yields comparable low values of $\chi_r^2$. At very
low values of $\chi_r^2$ ($<$0.1) the exponential law seems to provide a
slightly better fit, while at very high values of $\chi_r^2$ the
S\'ersic law is the one that more often yields a lower value for the
reduced $\chi_r^2$. In this sense, for the 21 exponential fits with
$\chi_r^2>1$ the $\chi_r^2$ of the corresponding S\'ersic fit is better
than that given by the exponential fit. However, this only represents
10\% of the sample. For only 5 of the profiles is $\chi_r^2$$>$3.

In some cases although the best fit yielded relatively low $\chi_r^2$
values, inspection of the profile showed that fit was obtained using
only a few points (never less than 5) with their errors strongly
dominated by the uncertainty in the background subtraction. In this
situation, the errors of the different points used for the fit are
largely correlated and the value of $\chi_r^2$ may be underestimated
(see Tables~\ref{table1} \& \ref{table2}). An example of this behavior
is seen at the $B$-band profile of HS~0029$+$1748.

The use of a S\'ersic law has the advantage of encompassing the
exponential law as a particular case. It also allows us to consider
for BCD galaxies a relationship between the S\'ersic-law index and
luminosity found for dwarf elliptical galaxies (see Graham \& Guzm\'an
2003). However, in the case of the USP of BCD galaxies, the region
available to us for the fitting is relatively small and it is located
at large distances from the galaxy center. Under these circumstances,
the uncertainties and degeneracies between the parameters of the
S\'ersic law become extremely large and highly dependent on the
particular region of the profile considered (Cair\'os et al$.$ 2003;
Noeske et al$.$ 2003; see also Section~\ref{results.sersicvsexp}).

\section{Results and discussion}
\label{results}

\subsection{Surface brightness and color profiles}
\label{results.sbp}

The surface brightness profiles shown in Figure~\ref{figure1} are
typically characterized by a high surface brightness component (HSB)
near the center of the galaxy superimposed on a nearly exponential
low-surface-brightness component (LSB) associated with its USP. These
characteristics in the profiles of BCD galaxies have been previously
observed by various authors (Loose \& Thuan 1986; Papaderos et al$.$
1996a,b; Cair\'os et al$.$ 2001a,b). Small departures of the surface
brightness profile of the USP from an exponential law have been
proposed. Doublier et al$.$ (1997, 1999) argued that in approximately
one fourth of the BCD galaxies considered by them the profiles were
better described by a de Vaucoleurs $r^{1/4}$ law. On the other hand,
Noeske et al$.$ (2003), using deep near-infrared imaging, have
recently proposed that a large fraction of BCD galaxies may have USP
with surface brightness profiles showing a central flattening similar
to the type-V profiles found by Binggeli \& Cameron (1991) in some
dwarf elliptical galaxies. In Section~\ref{results.lsb} we analyze in
detail the morphology of the USP in our BCD sample of galaxies.

The color profiles obtained indicate that the HSB component commonly
shows very blue colors, especially once the colors are corrected for
line and nebular-continuum contamination. At larger radii, where the
relative contribution of the HSB component becomes smaller, the colors
tend to get redder. In most of the galaxies the color profile flattens
at the radial position where the USP begins to dominate the galaxy's
global surface brightness profile. This behavior was first observed by
Papaderos et al$.$ (1996a,b) and more recently by Doublier et al$.$
(1997, 1999) and Cair\'os et al$.$ (2001a,b). Papaderos et al$.$
(1996a,b) called the position where the color profile flattens the
{\it transition radius}. Note that the much larger sample used in the
present work compared with previous studies allows us to obtain, for
the first time, statistically meaningful conclusions about the
structural properties of BCDs as a class of objects. Of the 104 BCD
galaxies in our sample, 70\% (72 objects) show this kind of flattening
(some examples are ISZ~399, Tol~2, Haro~2). About 17\% (18 galaxies)
show a progressive reddening of the color profile in the outer parts
of the galaxy. In some of the objects in this group the contamination
from the HSB component may still be important at faintest surface
brightness levels detected by our observations. Examples of these
objects are NGC~4861 and Haro~9. Six objects (5\%) show a bluing of
the color profile in the outermost parts of the galaxy. An example of
this type of behavior is seen in the profile of
Tol~1345$-$420. Finally, for a total of 9 galaxies the large errors in
the outermost ($B-R$) colors measured prevent us from determining the
degree of flattening of the color profile beyond the {\it transition
radius}.

\subsection{Underlying stellar population}
\label{results.lsb}

\subsubsection{S\'ersic vs$.$ exponential law}
\label{results.sersicvsexp}

The S\'ersic indices found by fitting both the $B$ and $R$-band
profiles suggest that the USP tends to have a surface
brightness profile somewhat steeper than an exponential law. Moreover,
about 40 of the galaxies show indices steeper than the de Vaucoleurs
profile ($n$=4). However, it is worth noting that (as we commented in
Section~\ref{analysis.fitting}) the values derived for the
S\'ersic-law parameters are highly uncertain and strongly dependent on
the region considered for the fit. In particular, they strongly depend
on the surface brightness of the few innermost points considered
during the fit. At radial distances close to, but beyond the position
of the {\it transition radius}, some profiles show surface
brightnesses in excess of what is expected from our best-fitting
exponential law. Since these are the innermost points considered in
our S\'ersic-law fit, their value has a critical impact on the
best-fitting index derived. We believe these
intermediate-radial-distance regions may well be contaminated by the
emission from the HSB component. This contamination would affect
substantially the shape of total surface brightness profile, but would
have a smaller impact on the color profile. Some examples are
NGC~1705, NGC~2915, Mrk~1423, NGC~3125, ESO~572-G025.

In order to understand the effect of the particular region considered
for the fit on the indices derived we have also computed the
best-fitting S\'ersic index within the region used for the fit to an
exponential law. Figure~\ref{figure2}g-i show the comparison between
the distributions of S\'ersic indices obtained in this way and those
obtained using different regions for the S\'ersic and exponential-law
fits. Although the number of objects with profiles steeper than
exponential is still significant, the number of objects with S\'ersic
indices $n$$>$4 is significantly lower when the fit is performed
within the same region used for the exponential fit.

The large differences obtained by the two methods demonstrate the
strong dependence of the best-fitting S\'ersic-law parameters on the
chosen set of points considered during the fitting procedure (see
Section~\ref{analysis.fitting}; see also Cair\'os et al$.$ 2003).
Therefore, in the rest of the paper we will use the best-fitting
exponential-law parameters derived in order to characterize the
structural properties of the USP in BCDs.  The use of very
deep images at near-infrared wavelengths should improve significantly
our knowledge about the detailed morphology of the USP in BCD galaxies
given the smaller contamination associated with the HSB component at
these wavelengths (see Noeske et al$.$ 2003 for a pilot study using a
sample of 12 BCDs).

\subsubsection{Structural properties, colors, and luminosity}
\label{results.therest}

The best-fit exponential and S\'ersic-law parameters for the galaxies
in the sample (including the $\chi_r^2$ of the best fit) are given in
Tables~\ref{table1} \& \ref{table2}, respectively. Table~\ref{table3}
shows the total, extrapolated $B$ and $R$ magnitudes of the
best-fitting exponential USP. We also provide the color (observed and
corrected for line and nebular-continuum contamination) of the USP
weight averaged over the region where the best-fitting exponential and
S\'ersic laws were obtained.

The scale-length distributions obtained for the $B$ and $R$ bands are
very similar, which confirms that the color gradients in the USP,
although present (see Section~\ref{results.sbp}), are not very
large. The mean values derived for the scale length of the USP are
$\log(\alpha)$=2.7$\pm$0.3 and 2.8$\pm$0.3, respectively for the $B$
and $R$-band profiles. With regard to the extrapolated central surface
brightness, the mean values obtained are
$\mu_{\mathrm{B,0}}$=21.7$\pm$1.3 and
$\mu_{\mathrm{R,0}}$=21.1$\pm$1.2\,mag\,arcsec$^{-2}$. For comparison,
Papaderos et al$.$ (1996b) obtained a slightly brighter average value
of $\mu_{\mathrm{B,0}}$=21.3\,mag\,arcsec$^{-2}$ for their sample of
12 BCD galaxies.

Figure~\ref{figure3} compares the integrated color of the galaxies in
the sample (see Paper~I) with the color of the USP (measured in the
region where the best-fitting exponential law was derived). In most of
the objects the ($B-R$) color of the envelope is redder by about
0.3\,mag than the integrated color, which has contributions from the
recent star formation. Only in those objects where the line and
nebular-continuum contamination is significant (e.g., UM~404, UM~417,
HS~0822$+$3542, I~Zw~18) is the integrated color redder than the color
of the USP (see Papaderos et al$.$ 2002 for a detailed study
of the impact of line and nebular-continuum contamination on the
observed broad-band properties of BCDs).

In Figure~\ref{figure4}a we compare the distribution of the galaxies
in our sample in the ($B-R$)-M$_{\mathrm{B,LSB}}$ diagram. The ($B-R$)
colors in this figure have been corrected for line and
nebular-continuum contamination, and the absolute magnitude plotted
(M$_{\mathrm{B,LSB}}$) refers to that of the best-fitting exponential
low-surface-brightness component associated with the USP. This plot
constitutes a direct means of comparing the properties of the USP of
BCD galaxies with those of other types of dwarf galaxies.

Since local dwarf elliptical galaxies (dE) show no significant star
formation and very shallow color gradients (e.g., Vader et al$.$ 1988)
we can directly compare this plot with the (integrated)
color-magnitude diagram of dE galaxies (circles in
Figure~\ref{figure4}a). Field dE galaxies (Parodi et al$.$ 2002), dE
in the Sculptor and Cen-A groups (Jerjen et al$.$ 2000) and cluster dE
in Virgo (Barazza, Binggeli, \& Jerjen 2003) and Perseus (Conselice,
Gallagher, \& Wyse 2003) are plotted. We also show the fiducial
color-magnitude relationship for the Coma cluster (solid line; Secker,
Harris, \& Plummer 1997). Finally, we have also included in this plot
a small sample of dwarf irregular galaxies for which colors and
luminosities of their USP are available (outlined stars; Parodi et
al$.$ 2002).

Figure~\ref{figure4}a shows that there are many BCD galaxies in our
sample which show colors and luminosities of their USP that are
comparable to those of elliptical galaxies. In Figure~\ref{figure4}b
the frequency histograms of the ($B-R$) color of dE galaxies (outline
histogram) and BCD galaxies (solid and hatched histograms) are
plotted. While there is significant overlap the USP of BCDs is (on
average) bluer than that of dE galaxies. However,
Figure~\ref{figure4}b also shows that the distribution of the color of
the USP of BCD galaxies with smooth envelopes (``E''-type;
nE and iE types according to Loose \& Thuan 1986) (solid histogram) is
more similar to that of dE galaxies than the one for BCD galaxies with
irregular envelopes (iI-type BCDs; hatched histogram).

These results suggest that a measurable fraction of the BCD galaxies
show USP with colors, luminosities, and apparently also morphologies
similar to those of dwarf ellipticals.

In order to quantitatively analyze the morphological differences
between BCD galaxies with red and blue USP we also compare the
extrapolated central surface brightness and scale lengths of the
galaxies in our sample (see Table~\ref{table1}) with those of BCD, dI,
and dE galaxies taken from the literature (P. Papaderos, private
communication; see also Papaderos et al$.$ 1996b). In
Figure~\ref{figure5}a we plot the extrapolated central surface
brightness against the luminosity (both in the $B$-band) of the
best-fitting exponential USP for the galaxies in our sample, and
compare them with those of dE (small dots), dI (stars), LSB (open
crosses) and other BCD galaxies from the literature (open
diamonds). Those BCDs in our sample offset by less than $\pm$0.4\,mag
from the color-magnitude relationship of dE galaxies are represented
by large dots. BCDs offset by more than this amount are shown as
filled diamonds. This allows us to separate BCD galaxies with red
envelopes from those with blue envelopes. The horizontal dotted line
at $\mu_{\mathrm{B,0}}$=22\,mag\,arcsec$^{-2}$ marks the separation
between BCDs and other types of dwarf galaxies, as proposed by
Papaderos et al$.$ (1996b). This figure shows that there are many BCD
galaxies with a central surface brightness of the USP fainter than
22\,mag\,arcsec$^{-2}$ and that in most of those galaxies this
component is as red as dE galaxies.

Figure~\ref{figure5}b shows the exponential scale length plotted
against luminosity for the same objects as in
Figure~\ref{figure5}a. Again, BCD galaxies with red envelopes show (on
average) scale lengths comparable to those seen in dE galaxies, and
larger than those of BCDs with blue envelopes. The dashed line shows
the least-squares fit to the distribution of dwarf elliptical galaxies
in this plot.

By analyzing the properties of the USP of the galaxies
individually we find that a total of 17 BCDs in our sample show
envelopes with (1) smooth elliptical morphologies (nE or iE types),
(2) dE-like colors, and (3) faint extrapolated central surface
brightnesses. This makes up slightly over 15\% of the whole sample.

\subsection{Implications on the unified evolutionary model of dwarf galaxies}
\label{results.dwarfs}

The fraction of BCDs with properties of their USP similar to those of
dwarf ellipticals may be even larger if, as proposed by Papaderos et
al$.$ (1996b), the structural properties of the USP of BCDs may vary
with time in response to changes in the gravitational potential driven
by the collective effect of stellar winds from massive stars and
supernova explosions (see next section for a discussion on this
topic).

The results presented above allow us to conclude that a significant
fraction of the BCDs in the nearby Universe (at least 15\%) are
consistent with being dwarf elliptical galaxies that are now
experiencing, or have recently experienced, an episode of active star
formation. These objects can be easily identified with the
``slowly-moving'', gas-accreting dE galaxies proposed by Silk, Wyse,
Shields (1987). According to this scenario, these galaxies are
expected to subsequently evolve into nucleated dwarf ellipticals,
following the sequence dE$\rightarrow$BCD$\rightarrow$dE,N. Note that
the possible evolution from BCDs to (nucleated) dE galaxies has been
recently questioned by Tajiri \& Kamaya (2002; see also van Zee et
al$.$ 2001) because of the difficulty in having these galaxies blow
away their HI envelopes.

\subsection{Impact of the recent star formation on the evolution of BCDs}
\label{results.sfh}

In this section we study the effect of the current star formation on
the structural properties of the USP of BCDs. According to Papaderos
et al$.$ (1996b) the offset between the scale length of dE and the USP
of some BCDs may be due to expansion of the USP in response to changes
in the gravitational field produced by collective supernova-driven
winds. Here we define the {\it degree of expansion} as the difference
(in logarithmic scale) between the scale length of the galaxy and the
average scale length of dwarf ellipticals of the same luminosity. The
latter quantity is obtained from the least-squares fit to the dE's
scale length and luminosity shown by the dashed line in
Figure~\ref{figure5}b. In Figure~\ref{figure6} the {\it degree of
expansion} is compared with the observed equivalent width of H$\alpha$
(panel {\bf a}) and with the difference between the galaxy color with
that of its USP (panel {\bf b}). Note that the errors in the {\it
degree of expansion} include the scatter in the least-squares fit used
to derive the average scale length of dwarf ellipticals at a given
luminosity.

Despite the strong dependence of the equivalent width of H$\alpha$ on
the age of the young stellar population for the case of instantaneous
star formation, if the expansion of the USP in BCDs is related to the
strength of the recent star-forming event we would expect to find a
correlation between EW(H$\alpha$) and the {\it degree of expansion}
(as defined above). This correlation should be even more evident if,
as it is thought, the recent star formation in BCDs takes place in
episodes of approximately constant star formation that last as long as
10$^{8}$\,yr (see Papaderos et al$.$ 2002 and references
therein). However, Figure~\ref{figure6}a does not show any obvious
correlation between EW(H$\alpha$) and the {\it degree of expansion},
which suggests that the current episode of star formation probably has
had little impact, if any, on the structural properties of the USP.

The ($B-R$) color is even more sensitive to the strength of the most
recent star formation episode than is the equivalent width of
H$\alpha$ (see Figure~6a of Paper~I). Thus, for a relatively evolved
USP, a recent episode of star formation with even a small burst
strength may have a strong impact on the observed ($B-R$) color of the
galaxy for a relatively long period of time.  However, in the case of
BCD galaxies with blue envelopes the difference in color between the
galaxy and its USP may be small even for relatively massive bursts.

Figure~\ref{figure6}b seems to shows a slight tendency for galaxies
with more negative {\it degree of expansion} to have slightly smaller
differences in color between the galaxy and its USP. This is opposite
to what we would expect if both the {\it degree of expansion} and the
difference in color measured would only depend on the burst strength
of the most recent star formation event.

In order to quantify the impact of this recent star formation on the
structural properties of the USP of BCDs we will use the formalism of
Papaderos et al$.$ (1996b). According to these authors we can write
the {\it degree of expansion} as
\begin{equation}
\log(\alpha) - <\log(\alpha_{\mathrm{dE}})> = \log\left(1- {{\mathfrak F_0} \over {1+\psi_{R_{\mathrm{H0}}}}}\right)
\label{equ1}
\end{equation}
where $\mathfrak F_0$ is the fraction of the visible mass ejected from
the galaxy as a consequence of the collective effect of
supernova-driven winds, and $\psi_{R_{\mathrm{H0}}}$ is the
dark-to-visible mass ratio inside the galaxy's Holmberg radius
($R_{\mathrm{H0}}$). Note that in our case $\mathfrak F_0$ is defined
to be positive. The simulations of the evolution of the ISM around
dwarf starburst galaxies carried out by MacLow \& Ferrara (1999)
indicate that $\mathfrak F_0$ is a strong function of the total
visible mass of the galaxy (M$_\mathrm{vis}$) and the kinetic energy
injection rate ($L_{\mathrm{kin}}$). We have used the $\mathfrak F_0$
values given by these authors for visible masses between 10$^6$ and
10$^9$\,M$_{\odot}$ and kinetic luminosities in the range
10$^{37}$-10$^{39}$\,erg\,s$^{-1}$. The fraction of mass ejected in
the case of kinetic luminosities $>$10$^{39}$\,erg\,s$^{-1}$ has been
determined using the following relationship, which adequately
reproduces (with an error $<$$\pm$0.12\,dex) the $\mathfrak F_0$
values given by MacLow \& Ferrara (1999),
\begin{equation}
\mathfrak F_0 \approx 10^{-10.84} \times \left({{\mathrm{M}_{\mathrm{vis}}} \over {\mathrm{M}_\odot}}\right)^{-1.7} \times \left({{L_{\mathrm{kin}}}\over{\mathrm{erg\,s}^{-1}}}\right)^{0.55}
\label{equ2}
\end{equation}
At very high values of $L_{\mathrm{kin}}$ and/or low values of
M$_{\mathrm{vis}}$, where this expression yields $\mathfrak F_0$$>$1,
the value of $\mathfrak F_0$ was set to 1. For the sake of simplicity
we assume the visible mass to be dominated by the galaxy's stellar
component. The kinetic luminosity per unit mass of a starburst remains
approximately constant for the first few 10$^{7}$\,yr and equal to
$\sim$10$^{35.5}$ erg\,s$^{-1}$\,M$_{\odot}^{-1}$ for a Salpeter IMF
and M$_{\mathrm{low}}$=1\,M$_{\odot}$ and
M$_{\mathrm{up}}$=100\,M$_{\odot}$ (Leitherer \& Heckman 1995). Thus,
Equation~\ref{equ1} can be written as
\begin{equation}
\log(\alpha) - <\log(\alpha_{\mathrm{dE}})> = \log\left(1- {{\min\left(1, 4.84 \times 10^8 \times b^{0.55} \times \left({{\mathrm{M}_*} \over {\mathrm{M}_\odot}}\right)^{-1.15}\right)} \over {1+\psi_{R_{\mathrm{H0}}}}}\right)
\label{equ3}
\end{equation}
where the burst strength $b$, is the ratio of the stellar mass of the
most recent episode of star formation to the galaxy's total stellar
mass. Finally, in order to derive the dependence of the color
difference [($B-R$)$-$($B-R$)$_{\mathrm{LSB}}$] on the burst strength
we used the predictions of evolutionary synthesis models (e.g$.$
Bruzual \& Charlot 2003). We adopted a 9-Gyr-old USP with
Z$_{\odot}$/5 metallicity and a burst of star formation with different
burst strengths in the range $b$=10$^{-4}$-1. For each value of the
burst strength we computed the color difference
[($B-R$)$-$($B-R$)$_{\mathrm{LSB}}$] averaged over the first 10\,Myr
of evolution of the burst.

Figure~\ref{figure6}b shows that in order for the BCD galaxies to show
the large {\it degree of expansion} measured, their stellar mass
should be smaller than 10$^{7}$\,M$_{\odot}$ and they should have
small dark-to-visible mass ratios. Even if the color of the USP were
to be much bluer than that of a 9-Gyr-old stellar population the vast
majority of the galaxies in our sample would be less massive than
10$^{8}$\,M$_{\odot}$ and the dark matter contribution within the
Holmberg radius would be negligible. This is required if the large
differences in scale length between BCDs and dE galaxies are to be
explained as due exclusively to the expansion of the underlying
stellar mass distribution of BCDs. These conditions are certainly not
fulfilled by the galaxies in our sample, where the stellar masses can
be much larger than this number (M$_{K}$ for our sample can be as high
as $-$21\,mag ; Paper~I) and where significant amounts of dark matter
are thought to be present (Ferrara \& Tolstoy 2000). This result again
argues against the current star formation having a strong impact on
the structural properties of the USP of BCDs. However,
detailed studies of individual objects are required to confirm this in
all cases.

\section{Conclusions}
\label{conclusions}

In summary,

\begin{itemize}

\item We have presented the surface brightness profiles in $B$, $R$,
and H$\alpha$, for a total of 114 galaxies taken from the Palomar/Las
Campanas Imaging Atlas of BCD galaxies. A total 104 of the galaxies
are classified as BCDs (see Paper~I). The profiles in the continuum
bands are characterized by the presence of a HSB component on top of
the nearly exponential low-surface-brightness component associated
with the galaxy's underlying stellar population (USP). At large
galactocentric radii the color profiles of 70\% of the galaxies
flatten. This flattening occurs approximately at the position where
the USP starts to dominate the galaxy surface brightness profile.

\item The color of the USP (corrected for line and
nebular-continuum emission) is systematically redder than the observed
integrated color. The color of the USP is bluer than the integrated
one only in those objects with the highest equivalent widths of
H$\alpha$ (where the line and nebular-continuum emission is
significant; i.e$.$ EW(H$\alpha$) larger that a few hundred
angstroms).

\item We find that galaxies with relatively red USP (($B-R$)$\geq$1\,mag) 
show structural properties compatible with those of dwarf elliptical
galaxies. They show smoother (continuum) light distributions, fainter
extrapolated central surface brightness, and larger scale lengths than
BCD galaxies with blue envelopes. This result indicates that a
non-negligible fraction of the BCD galaxies ($\sim$15\%) could be
dwarf ellipticals that are now experiencing (or have recently
experienced) an episode of star formation.

\item We do not find any correlation between the equivalent width of
H$\alpha$ and the {\it degree of expansion} of the USP, the latter
being defined as the difference in scale length between BCD and dE
galaxies of identical luminosity. The difference measured between the
scale length of BCDs and dE galaxies is much larger than that expected
from changes in the gravitational potential due to the collective
effect of supernova-driven winds, especially considering the
relatively small differences in ($B-R$) color between the galaxies and
their USP. This suggests that the level of recent star formation in
BCDs does not have a significant impact on the structural properties
of these galaxies. A detailed study of a large number of individual
objects is needed to confirm this.

\end{itemize}

\acknowledgments We are grateful to the Palomar and Las Campanas
Observatories staff for their support and hospitality, and to the
Caltech/Palomar and OCIW/Las Campanas Time Allocation Committees for
the generous allocation of time to this project. AGdP acknowledges
financial support from the GALEX mission. AGdP is also partially
supported by the CONACYT (Mexico) grant 36132-E, the Spanish Programa
Nacional de Astronom\'{\i}a y Astrof\'{\i}sica under grant
AYA2000-1790, and by NASA through grant HST-AR-10321 from STScI. This
research has made use of the NASA/IPAC Extragalactic Database (NED)
which is operated by the Jet Propulsion Laboratory, California
Institute of Technology, under contract with the National Aeronautics
and Space Administration. We would like also to thank K. G. Noeske,
C. S\'{a}nchez Contreras, and S. Boissier for valuable discussions and
to P. Papaderos for providing his compilation of structural properties
of dwarf galaxies. We are grateful to the anonymous referee for
her/his helpful comments and suggestions.







\clearpage
\begin{deluxetable}{lrrrrrrc}
\tabletypesize{\small}
\tablecaption{Exponential-law fitting parameters\label{table1}}
\tablewidth{0pt}
\tablehead{
\colhead{Object Name} & 
\colhead{$\alpha_B$} & 
\colhead{$\mu_{B,0}$} & 
\colhead{$\chi_r^2$}  & 
\colhead{$\alpha_R$} & 
\colhead{$\mu_{R,0}$} & 
\colhead{$\chi_r^2$}  & 
\colhead{Notes} 
\\
\colhead{} & 
\colhead{(arcsec)} & 
\colhead{(mag/\sq)} & 
\colhead{} & 
\colhead{(arcsec)} & 
\colhead{(mag/\sq)} & 
\colhead{} & 
\colhead{} 
\\
\colhead{(1)} & 
\colhead{(2)} & 
\colhead{(3)} & 
\colhead{(4)} & 
\colhead{(5)} & 
\colhead{(6)} & 
\colhead{(7)} & 
\colhead{(8)} 
}
\startdata
IC~10           & 100.8$\pm$3.8 &  17.48$\pm$0.06 &  0.17 & 112.1$\pm$1.8 &  17.48$\pm$0.09 &  0.00 &      \\
HS~0029+1748    &   5.5$\pm$0.4 &  23.09$\pm$0.18 &  0.02 &   3.8$\pm$0.2 &  21.14$\pm$0.17 &  0.06 &  (a) \\
Haro~14         &   6.1$\pm$0.1 &  19.36$\pm$0.10 &  0.49 &   6.7$\pm$0.1 &  18.98$\pm$0.15 &  0.05 &      \\
UM~285          &   2.6$\pm$0.2 &  21.86$\pm$0.24 &  0.32 &   3.7$\pm$0.4 &  22.85$\pm$0.29 &  0.36 &  (a) \\
UCM~0049-0045   &   8.7$\pm$0.7 &  22.46$\pm$0.16 &  0.81 &  10.6$\pm$0.4 &  21.53$\pm$0.27 &  0.09 &      \\
UM~323          &   2.3$\pm$0.1 &  19.58$\pm$0.17 &  0.30 &   2.4$\pm$0.1 &  19.11$\pm$0.14 &  0.86 &      \\
Mrk~996         &   4.2$\pm$0.1 &  20.56$\pm$0.08 &  0.30 &   4.5$\pm$0.1 &  19.75$\pm$0.15 &  1.06 &      \\
UM~133          &   9.9$\pm$0.4 &  23.03$\pm$0.13 &  0.29 &  12.0$\pm$0.6 &  22.37$\pm$0.18 &  0.15 &      \\
\enddata
\tablecomments{Columns stand for: (1) Galaxy name. (2) Scale length of
the best-fitting exponential profile of the USP in the $B$-band
(arcsec). (3) Extrapolated central surface brightness of the USP in
the $B$-band (mag/arcsec$^2$). (4) Reduced $\chi^2$ of the best
exponential fit in the $B$-band.  (5-7) The same as (2-4) for the
$R$-band. (8) Notes: (a) Errors in the points used for the fit are
correlated. $\chi_r^2$ could be underestimated. (b) Errors strongly
correlated. The value of $\chi_r^2$ should be considered a lower limit
(see text for details).}
\end{deluxetable}

\clearpage
\begin{deluxetable}{lrrrrrrrrrrrrc}
\tabletypesize{\tiny}
\rotate
\tablecaption{S\'ersic-law fitting parameters\label{table2}}
\tablewidth{0pt}
\tablehead{
\colhead{Object Name} & 
\colhead{$1/n_B$} & 
\colhead{$r_{B,0}$} & 
\colhead{$\mu_{B,0}$} & 
\colhead{$\chi_r^2$}  & 
\colhead{$1/n_R$} & 
\colhead{$r_{R,0}$} & 
\colhead{$\mu_{R,0}$} & 
\colhead{$\chi_r^2$}  & 
\colhead{$1/n_{\mathrm{H}\alpha}$} & 
\colhead{$r_{\mathrm{H}\alpha,0}$} & 
\colhead{log $f_{\mathrm{H}\alpha,0}$} & 
\colhead{$\chi_r^2$} & 
\colhead{Notes} 
\\
\colhead{} & 
\colhead{} & 
\colhead{(arcsec)} & 
\colhead{(mag/\sq)} & 
\colhead{} & 
\colhead{} & 
\colhead{(arcsec)} & 
\colhead{(mag/\sq)} & 
\colhead{} & 
\colhead{} & 
\colhead{(arcsec)} & 
\colhead{(cgs/\sq)} & 
\colhead{} & 
\colhead{} 
}
\startdata
IC~10           & 1.11$\pm$0.53 &         115$\pm$60 &     17.64$\pm$5.67 &  0.17 & 0.70$\pm$1.43 &     50$\pm$100 &     16.55$^{\mathrm{c}}$ &  0.00 & 0.41$\pm$0.10 &      7.02$\pm$ 8.95 &    -11.31$\pm$1.01 &  5.97 &      \\
HS~0029+1748    & 0.33$\pm$0.12 &      0.01$\pm$0.15 &     15.48$\pm$7.66 &  0.21 & 0.37$\pm$0.29 &      0.06$\pm$1.02 &     17.01$^{\mathrm{c}}$ &  0.05 & 0.05$\pm$0.01 & $<$0.01 & \nodata   &  1.29 &      \\
Haro~14         & 0.18$\pm$0.22 & $<$0.01 & \nodata            &  0.13 & 0.50$\pm$0.23 &      0.33$\pm$1.63 &     13.48$^{\mathrm{c}}$ &  0.07 & 1.19$\pm$0.01 &     10.45$\pm$0.04 &    -14.55$^{\mathrm{c}}$ & $>$100 &  (a) \\
UM~285          & 0.05$\pm$0.05 & $<$0.01 & \nodata            &  0.56 & 0.05$\pm$0.00 & $<$0.01 & \nodata &  0.77 & 1.14$\pm$0.69 &      4.28$\pm$2.55 &    -14.60$^{\mathrm{c}}$ &  0.32 &      \\
UCM~0049-0045   & 0.05$\pm$0.00 & $<$0.01 & \nodata &  0.59 & 0.06$\pm$0.03 & $<$0.01 & \nodata            &  0.40 & 2.91$\pm$0.21 &      9.95$\pm$0.26 &    -14.83$\pm$0.08 &  0.21 &      \\
UM~323          & 0.45$\pm$0.27 &      0.07$\pm$0.63 &     14.06$^{\mathrm{c}}$ &  0.12 & 0.52$\pm$0.19 &      0.21$\pm$0.49 &     15.45$\pm$8.78 &  0.23 & 0.05$\pm$0.17 & $<$0.01 & \nodata   &  0.80 &      \\
Mrk~996         & 0.67$\pm$0.11 &      1.31$\pm$0.76 &     18.91$\pm$0.89 &  0.17 & 0.66$\pm$0.05 &      1.36$\pm$0.38 &     18.12$\pm$0.38 &  0.35 & 0.05$\pm$0.00 & $<$0.01 & \nodata   &  1.22 &      \\
UM~133          & 0.56$\pm$0.66 &      1.66$\pm$8.06 &     20.84$^{\mathrm{c}}$ &  0.28 & 0.05$\pm$0.41 & $<$0.01 & \nodata            &  0.13 & 0.60$\pm$0.08 &      4.02$\pm$1.25 &    -14.03$\pm$0.24 &  2.71 &      \\
\enddata
\tablecomments{Columns stand for: (1) Galaxy name. (2) Index of the
best-fitting S\'ersic profile for the USP in the $B$-band. (3) Scale
length of the best-fitting S\'ersic profile in the $B$-band
(arcsec). (4) Extrapolated central surface brightness of the
best-fitting S\'ersic profile in the $B$-band (mag/arcsec$^2$). (5)
Reduced $\chi^2$ of the best S\'ersic profile fit in the $B$-band.
(10-13) The same as (6-9) for the $R$-band. (14) Notes: (a) Errors in
the points used for the fit are correlated. $\chi_r^2$ could be
underestimated. (b) Errors strongly correlated. The value of
$\chi_r^2$ should be considered a lower limit (see text for
details). (c) These quantities have uncertainties larger than
10\,mag\,arcsec$^{-2}$ (10\,dex) in $\mu_{B,0}$ or $\mu_{R,0}$ (log
$f_{\mathrm{H}\alpha,0}$). In those cases where the best S\'ersic fit
yielded (unrealistic) negative values for the $B$ and $R$-bands
extrapolated central surface brightness no data are given.}
\end{deluxetable}

\clearpage
\begin{deluxetable}{lrrrrrr}
\tabletypesize{\small}
\tablecaption{Properties of the Underlying Stellar Population\label{table3}}
\tablewidth{0pt}
\tablehead{
\colhead{Object Name} & 
\colhead{$B_{\mathrm{LSB},\infty}^{\mathrm{E}}$} & 
\colhead{$R_{\mathrm{LSB},\infty}^{\mathrm{E}}$} & 
\colhead{(B-R)$_{\mathrm{LSB}}^{\mathrm{E}}$} & 
\colhead{(B-R)$_{\mathrm{LSB}}^{\mathrm{E,c}}$} & 
\colhead{(B-R)$_{\mathrm{LSB}}^{\mathrm{S}}$} & 
\colhead{(B-R)$_{\mathrm{LSB}}^{\mathrm{S,c}}$} 
\\
\colhead{} & 
\colhead{(mag)} & 
\colhead{(mag)} & 
\colhead{(mag)} & 
\colhead{(mag)} & 
\colhead{(mag)} & 
\colhead{(mag)}   
}
\startdata
IC~10           &   5.47$\pm$0.10 &   5.24$\pm$0.09 &   0.23$\pm$0.14 &   0.22$\pm$0.12 &   0.23$\pm$0.14 &   0.22$\pm$0.12 \\
HS~0029+1748    &  17.38$\pm$0.24 &  16.27$\pm$0.19 &   0.66$\pm$0.16 &   0.65$\pm$0.15 &   0.56$\pm$0.15 &   0.49$\pm$0.15 \\
Haro~14         &  13.44$\pm$0.11 &  12.87$\pm$0.15 &   0.90$\pm$0.15 &   0.89$\pm$0.15 &   0.85$\pm$0.15 &   0.85$\pm$0.15 \\
UM~285          &  17.77$\pm$0.28 &  18.04$\pm$0.36 &   0.12$\pm$0.11 &   0.14$\pm$0.08 &   0.09$\pm$0.08 &   0.14$\pm$0.08 \\
UCM~0049-0045   &  15.76$\pm$0.23 &  14.42$\pm$0.28 &   1.47$\pm$0.27 &   1.40$\pm$0.25 &   1.77$\pm$0.25 &   1.78$\pm$0.25 \\
UM~323          &  15.81$\pm$0.19 &  15.17$\pm$0.16 &   0.73$\pm$0.09 &   0.70$\pm$0.09 &   0.73$\pm$0.09 &   0.70$\pm$0.09 \\
Mrk~996         &  15.46$\pm$0.09 &  14.51$\pm$0.16 &   1.00$\pm$0.14 &   1.00$\pm$0.14 &   0.98$\pm$0.14 &   0.98$\pm$0.14 \\
UM~133          &  16.06$\pm$0.16 &  14.97$\pm$0.20 &   1.17$\pm$0.15 &   1.40$\pm$0.14 &   1.16$\pm$0.15 &   1.40$\pm$0.14 \\
\enddata
\tablecomments{Columns stand for: (1) Galaxy name. (2) Total $B$-band
magnitude of the best-fitting exponential profile fitted to the
USP surface brightness distribution
(corrected for Galactic extinction). (3) The same as (2) for the
$R$-band. (4) ($B$$-$$R$) color of the underling stellar population in
the region where the best-fitting exponential profile was derived. (5)
The same as (4) corrected for line and nebular-continuum
contamination. (6) ($B$$-$$R$) color of the underling stellar
population in the region where the best-fitting S\'ersic profile was
derived. (7) The same as (6) corrected for line and nebular-continuum
contamination.}
\end{deluxetable}

\clearpage
\begin{figure}
\figurenum{1}
\Large{postscript files of panels f1a-f1o of figure 1 are available online at http://www.ociw.edu/\textasciitilde agpaz/astro-ph/apjs2004/\\}
\caption{Surface brightness and color profiles of the galaxies in the
BCD sample. {\bf Left panel:} For each galaxy we plot the two surface
brightness profiles (along with the 1-$\sigma$ errors), one each for
the $B$ (asterisks) and $R$ band (filled circles) data. The error bars
at the bottom-left corner of the diagram show the error associated
with the flux calibration of the $B$ (blue) and $R$-band (red)
images. We also plot the best-fitting exponential (solid blue for the
$B$-band and solid red lines for $R$) and S\'ersic (magenta lines for
$B$ and orange for $R$) profiles of the USP, and the corresponding
extrapolation to the center (dashed lines). The vertical ticks at the
bottom of the plot indicate the position outward from which the
surface brightness profile of the USP was fitted. In those galaxies
where the inner radius plotted is smaller than the image HWHM,
horizontal ticks at the left side of the plot show the extension of
the $B$ (blue) and $R$-bands (red) PSF HWHM. {\bf Top-right panel:} We
show the ($B-R$) observed (black points) and ionized-gas
contamination-corrected (gray points) color profiles with their
corresponding 1-$\sigma$ errors. The horizontal red (orange) line
marks the average ($B-R$) color of the USP and the region where the
best-fitting exponential (S\'ersic) profile was derived. The error bar
at the bottom-right corner of this diagram shows the error in the
($B-R$) color due to flux calibration uncertainties. {\bf Bottom-right
panel:} H$\alpha$ surface brightness profile in cgs units
(erg\,s$^{-1}$\,cm$^{-2}$\,arcsec$^{-2}$) and their corresponding
1-$\sigma$ errors. The green line shows the best-fitting S\'ersic
profile.
\label{figure1}}
\end{figure}

\clearpage
\begin{figure}
\figurenum{2}
\plotone{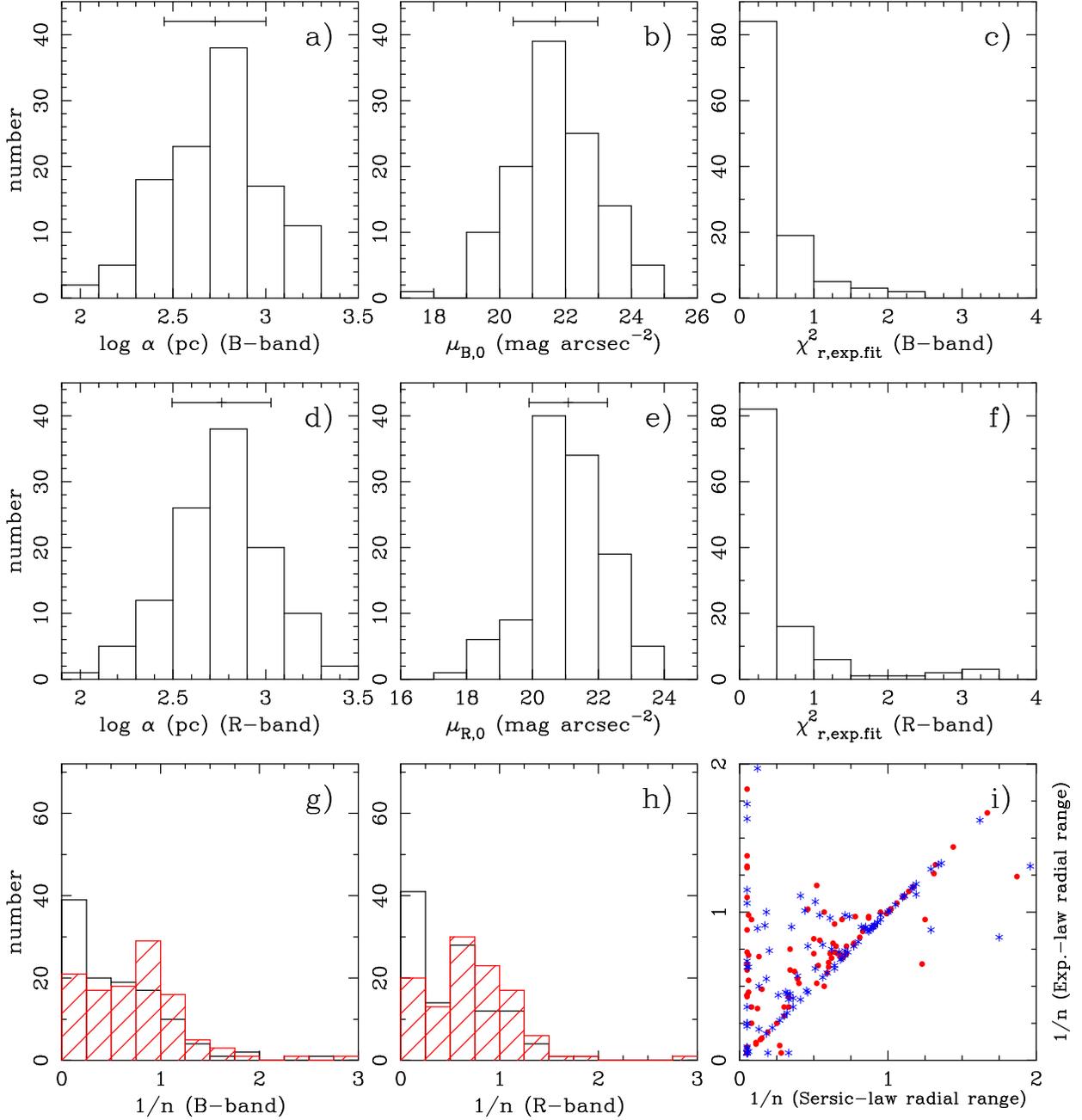}
\caption{Histograms of the properties derived from the galaxies
surface brightness profiles. {\bf a)} Scale length of the best-fitting
exponential profile of the USP in the $B$-band. {\bf b)} Extrapolated
central surface brightness of the best-fitting exponential profile in
the $B$-band. {\bf c)} Reduced $\chi^2$ of the best-fitting
exponential profile in the $B$-band. {\bf d,e,f)} The same as {\bf
a,b,c} for the $R$-band. {\bf g)} Best-fitting S\'ersic index for the USP in the
$B$-band. Hatched histograms represent the best-fitting S\'ersic
indices obtained using the same points in the profile that for the
exponential-law fit. {\bf h)} The same as {\bf g} for the $R$
band. {\bf i)} Comparison between the S\'ersic indices obtained using
the radial ranges derived from the exponential-law and S\'ersic-law
fits for the $B$ (asterisks) and $R$ (filled circles) bands. The tick
marks shown in panels {\bf a,b,d,e} represent the mean and
mean$\pm$1-$\sigma$ values of the corresponding
distribution.\label{figure2}}
\end{figure}

\clearpage
\begin{figure}
\figurenum{3}
\plotone{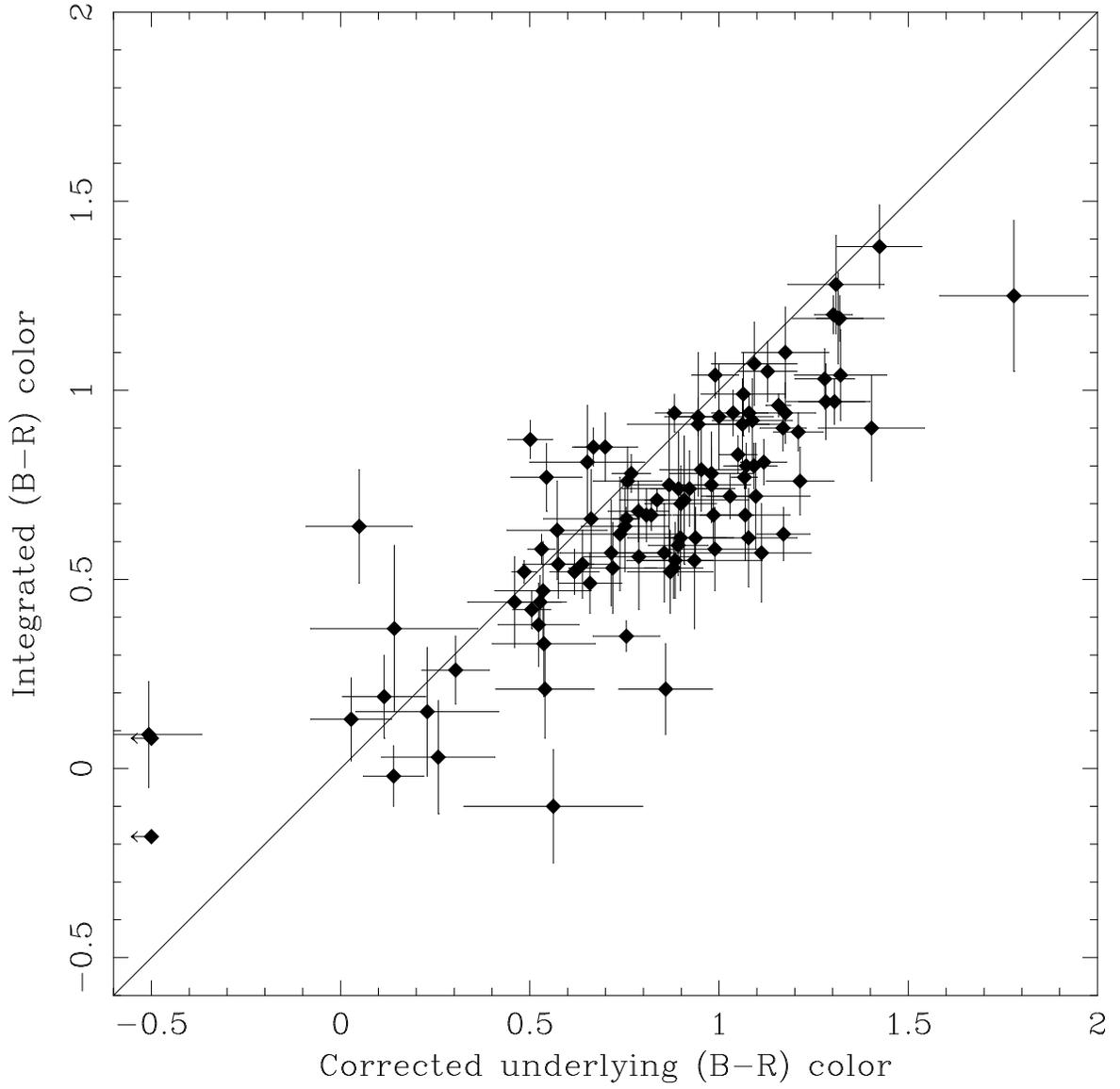}
\caption{Comparison between the integrated ($B-R$) color (from Paper~I) of the galaxies and the color of their USP.\label{figure3}}
\end{figure}

\clearpage
\begin{figure}
\figurenum{4}
\plotone{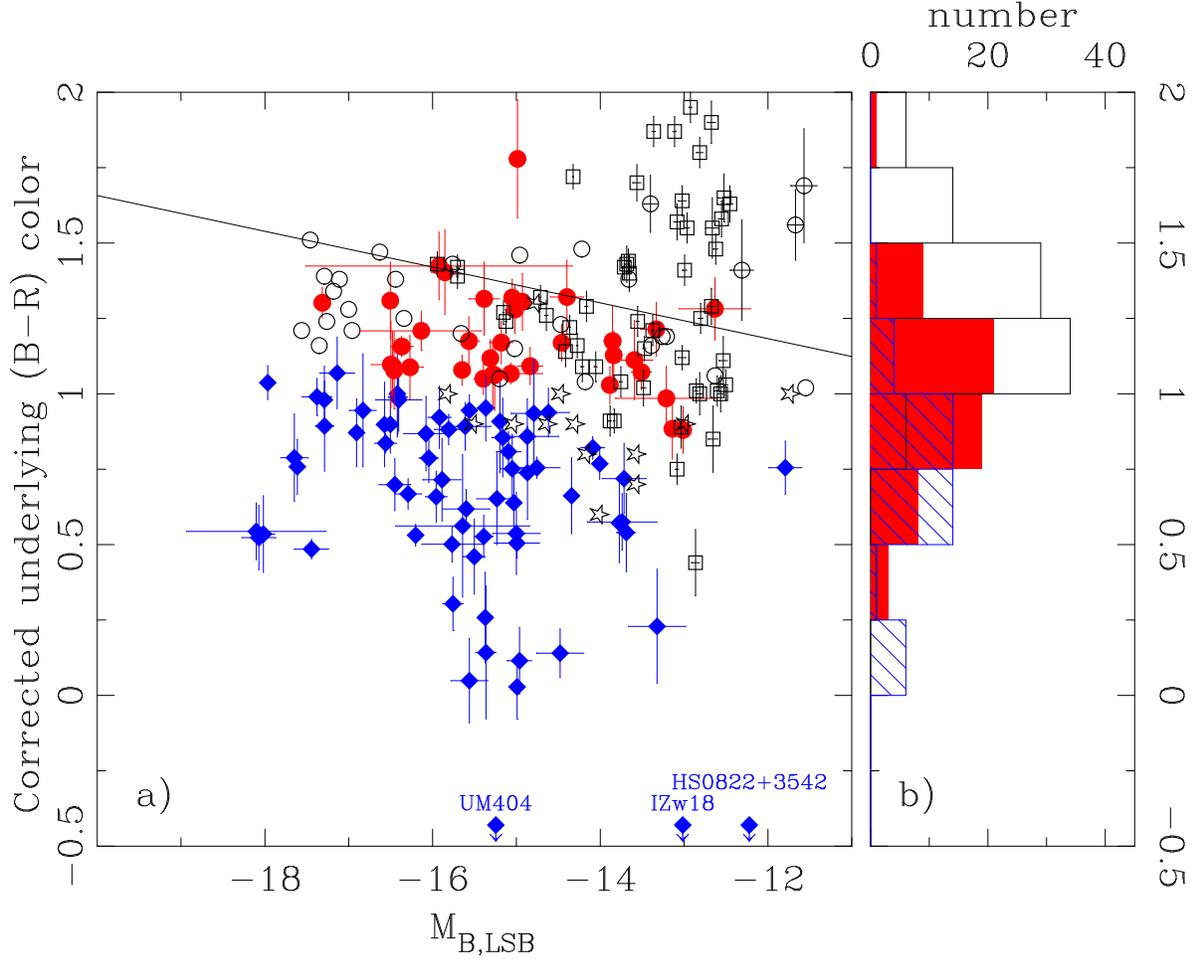}
\caption{{\bf a)} Comparison between the ($B-R$) color of the
USP and its absolute magnitude computed from the best exponential fit
to the surface-brightness profile of the USP. BCD galaxies in our
sample having blue (red) envelopes are represented by filled diamonds
(dots) (see text for more details). Dwarf elliptical (open circles and
squares) and dwarf irregular galaxies (stars) from the literature are
also plotted. The solid line represents the color-magnitude
relationship of dwarf elliptical galaxies in the Coma cluster (Secker
et al$.$ 1997). Note that the galaxies from Conselice et al$.$ (2003)
(open squares) were classified as dwarf ellipticals based exclusively
on their structural properties. {\bf b)} Histogram of the ($B-R$)
color of the USP for ``E''-type (filled histogram) and ``I''-type
(cross-hatched histogram) BCDs in our sample and the reference sample
of dwarf ellipticals (outlined histogram).
\label{figure4}}
\end{figure}

\clearpage
\begin{figure}
\figurenum{5}
\plotone{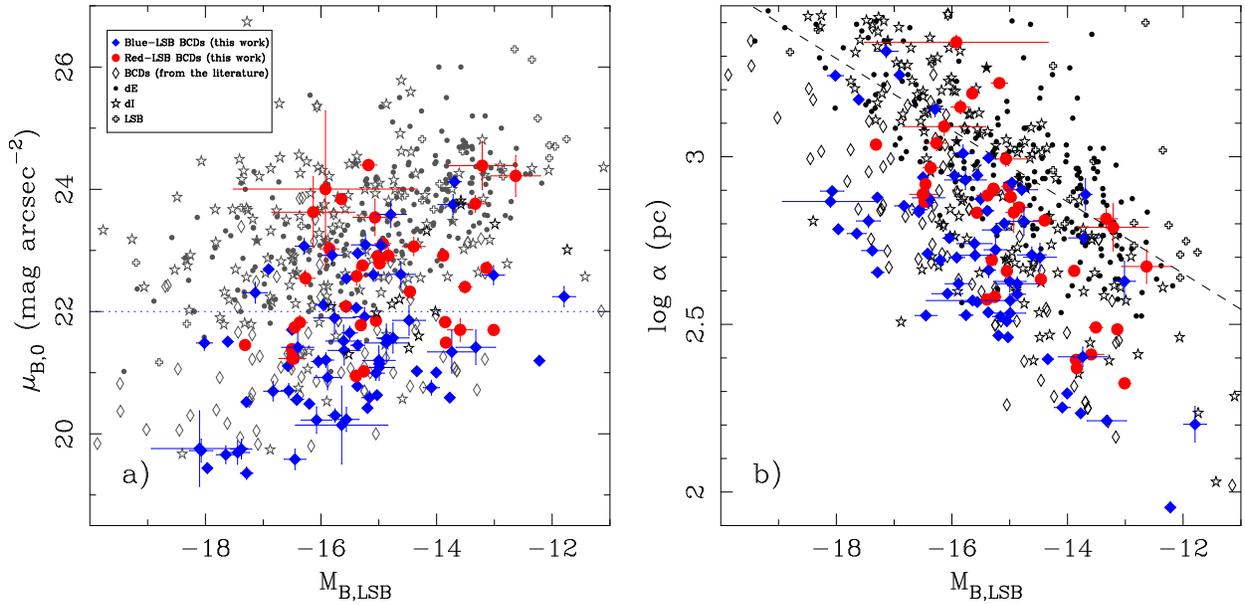}
\caption{Correlations between the structural properties of the
USP and luminosity. BCD galaxies in our
sample having blue (red) envelopes are represented by filled diamonds
(dots) (see text for more details). The properties of dE (small dots),
dI (stars), LSB (open crosses) and other BCD galaxies from the
literature (open diamonds) are also shown.\label{figure5}}
\end{figure}

\clearpage
\begin{figure}
\figurenum{6}
\epsscale{0.6}
\plotone{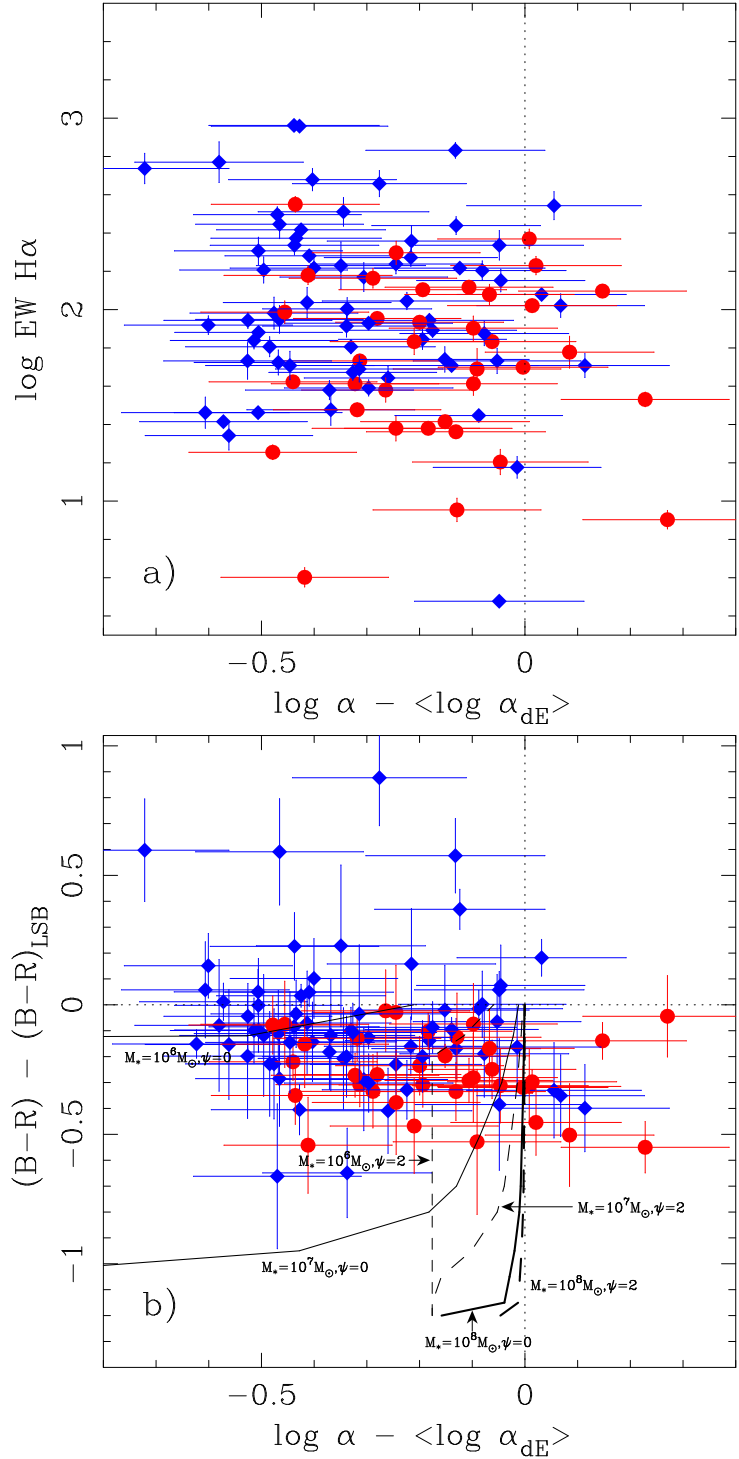}
\caption{{\bf a)} H$\alpha$ equivalent width $vs.$ the {\it degree of
expansion} (offset between the scale length of a BCD galaxy and the
average scale length of dwarf ellipticals of the same
luminosity). Different symbols are used for galaxies with blue
(diamonds) and red (dots) envelopes. {\bf b)} Difference in color
between the galaxies and their USP compared
with the {\it degree of expansion} of the USP. Models for different stellar masses and dark-to-visible
mass ratios are also shown (see text).\label{figure6}}
\end{figure}

\end{document}